\documentclass[fleqn,twoside,twocolumn,nofootinbib]{revtex4} 
\usepackage[sec]{ujp} 
\begin{document}
\title[STRESS-MAGNETIC FIELD PHASE DIAGRAM OF MULTIFERROICS]
{ON THE THEORY OF STRESS-MAGNETIC FIELD PHASE DIAGRAM OF THE FINITE SIZE MULTIFERROICS:
 COMPETITION BETWEEN FERRO- AND ANTIFERROMAGNETIC DOMAINS}
\author{H.V. GOMONAY}
\affiliation{Bogolyubov Institute for Theoretical Physics, Nat. Acad. of Sci. of Ukraine}
\address{14b, Metrolohichna Str., Kyiv 03143, Ukraine}
\email{malyshen@ukrpack.net}
\affiliation{National Technical
University of Ukraine ``KPI''}
\address{37, Peremogy Ave., Kyiv 03056, Ukraine}
\author{I.G. KORNIIENKO}
\affiliation{National Technical University of Ukraine ``KPI''}
\address{37, Peremogy Ave., Kyiv 03056, Ukraine}
\author{V.M. LOKTEV}
\affiliation{Bogolyubov Institute for Theoretical Physics, Nat. Acad. of Sci. of Ukraine}
\address{14b, Metrolohichna Str., Kyiv 03143, Ukraine}
\udk{537.9, 537.622,\\[-3pt] 621.318.14-416}
 \pacs{75.85.+t, 75.60.Ch,\\[-3pt] 46.25.Hf, 75.50.Ee}
\razd{\secvii}

\setcounter{page}{659}%
\maketitle
\begin{abstract}
Macroscopic properties of multiferroics, the systems that show
simultaneously two types of ordering, could be controlled by the
external fields of different nature. We analyze the behavior of
multiferroics with antiferro-(AFM) and ferromagnetic (FM) ordering
under the action of external magnetic and stress fields. A
combination of these two fields makes it possible to achieve
macroscopic states with different domain structures. The two-domain
state obtained in this way  shows a linear dependence of macroscopic
strain vs magnetic field which is unusual for AFMs. A small but
nonzero stress applied to the sample can also result in the bias of
the magnetization vs magnetic field dependence.
\end{abstract}

\section{Introduction}

Present-day technologies of information storage are based on the
\emph{i}) combination of the materials with different magnetic,
electronic, elastic properties; \emph{ii}) step-like variation of
the magnetic, electronic, elastic properties within a small,
nano-sized sample; \emph{iii}) combined application of different
fields to implement the control over macroscopic properties. The
first tendency initiated the growing interest in the natural and
synthetic multiferroics, i.e., materials that show simultaneously at
least two different order parameters (like the magnetic and electric
polarization). The second tendency is related to the appearance of a
giant variation (as compared with that in homogeneous samples) of
macroscopic properties like the magnetoresistance and, as a
consequence, the high susceptibility of samples to external fields.
The last approach opens a way to produce macroscopic states
unreacheable in other ways or the desired states much more
effectively. It should be noted that the sensitivity of
multiferroics to two fields of different nature instead of one
should be much greater than that of the
single-ferroics\footnote{Single-ferroics could also be controlled by
two different fields in the case where they show some
cross-correlated effects like the magnetostriction or the
piezoeffect.}. Really,  in  multiferroics, two order parameters have
different origin and different  symmetry  and  even  may  appear at
different temperatures, while the second-order parameter in
single-ferroics, which is also the secondary  one, contributes to
the values of characteristic fields and susceptibilities,  but
gives rise  to no new qualitative effects. So, we can expect that a
combination of two fields in multiferroics results in the variety of
states with different symmetries that could be used for the
effective information storage.

Many multiferroics show strong
magnetoelastic effects along with ferro- or antiferromagnetic
(FM and AFM, respectively) and ferroelectric ordering (see, e.g., [1--5]). In this case, the macroscopic
properties could be controlled by the simultaneous application of stress
and magnetic/electric fields. An external stress can be used for
the compensation of internal strains that produce the undesirable
backswitching of a ferroelectric polarization in some multiferroics
(like BiFeO$_3$) [4]. Additional stresses could be also produced in
the course of epitaxial growth (on a substrate with properly
chosen misfit), like it was done, e.g.,  in Ref. [6]. Such a
``plugged'' stress can stabilize a certain type of domain
structure (e.g., by eliminating undesirable domains) and thus
facilitate the control over macroscopic properties by external
magnetic/electric fields.

A combination of stress and magnetic/electric fields is also
interesting from the fundamental point of view, because these fields
have not only different symmetries, but also different tensor nature.
Here, we study the behavior of a multidomain magnetic
system in the presence of stress and magnetic fields.  A good example
of such a system can be found among materials that show
simultaneously the FM and AFM orderings in various systems of
magnetic ions (see, e.g., Mn-based antiperovskites [7, 8] and a high-temperature
superconductor Sr$_2$Cu$_3$O$_4$Cl$_2$ [9]) or in the
different parts of a sample (like manganite
Pr$_{0.5}$Ca$_{0.5}$MnO$_3$ [2, 3], in which the FM and AFM phases
coexist in a wide range of the temperature and the field). It is also
important that an AFM subsystem is usually a source of the
pronounced magnetostrictive strain and thus interacts with a stress,
while the FM subsystem is strongly coupled with an external magnetic
field. Moreover, in the finite-size samples, the AFM-induced
magnetostriction and the FM magnetization are the sources of different
shape-induced (destressing and demagnetizing) effects that compete
with external fields and may also compete with each other.

The main question which we address below is
whether a combination of magnetic and stress fields may reveal some
new features and facilitate the control over macroscopic properties
of a sample. We will analyze the phase diagrams of the homogeneous and
multidomain states and the field dependence of the macroscopic strain and
the magnetization. We argue that the presence of the FM component gives rise
to a substantial increase in the strain susceptibility to the
external magnetic field, and the application of a stress can induce the
magnetic bias seen as a shift of the magnetization curve.

\section{Combined Influence of a Magnetic Field and a Stress on the Magnetic System: Qualitative
Considerations}

 Let us consider a system that shows simultaneously
FM and AFM orderings, no matter how it is realized -- as a mixture of
phases (like in perovskite manganites [2, 3]) or as a FM+AFM
multiferroic (like Sr$_{2}$Cu$_{3}$O$_{4}$Cl$_{2}$ discussed in
details below). Suppose further that the system shows a noticeable
magnetoelastic effect which originates from the coupling between the AFM
order parameter and the lattice. Which mechanisms are responsible for
the equilibrium domain structure (DS) in this case, and how this DS could
be controlled? It is well known that the fraction of FM domains (in
a pure FM [10] or in a mixture with the AFM phase [11, 12]) is governed by
magnetostatic effects usually described with the stray energy
\begin{equation}\label{demag_1}
 \Phi_{\rm stray}=\frac{V}{2}N_{jk}^{\rm dm}\langle M^{\rm{macro}}_jM^{\rm{macro}}_k\rangle,
\end{equation}
where $\mathbf{M}^{\rm{macro}}$ is the macroscopic magnetization,
brackets $\langle \dots  \rangle$ mean the averaging over the sample
volume $V,$ and the second-rank demagnetization  tensor  $N_{jk}^{\rm
dm}$ depends upon the sample shape.

On the contrary, the AFM vectors themselves do not produce any
magnetostatic field. However, due to magnetostriction, an AFM
component produces the so-called destressing fields of elastic
nature [13] that give rise   to   the  formation  of  an
equilibrium  DS.

In the case of an FM+AFM phase mixture, the inclusions of a new (FM or
AFM, depending upon the direction of a phase transition) phase could
be considered as elastic dipoles (due to the strain misfit
between the FM and AFM phases). In the case of AFM+FM multiferroic, the
analogous dipoles arise at the sample surface (see [14] for
details). In both cases, the elastic dipoles produce long-range
fields that could be taken into account phenomenologically with the
use of the destressing energy
\begin{equation}\label{destress_1}
 \Phi_{\rm dest}=\frac{V}{2}N_{jklm}^{\rm des}\langle L_jL_k\rangle\langle L_lL_m\rangle,
\end{equation}
where $\mathbf {L}$ is an AFM order parameter, and the 4-th rank
destressing tensor $N_{jklm}^{\rm des}$, like $N_{jk}^{\rm dm}$,
depends upon the shape of the sample or/and inclusions and upon
the magnetoelastic coupling strength.

Both the stray (Eq. (1)) and destressing (Eq. (2)) contributions to
the free energy of a sample constrain a variation of the domain
fraction in the presence of external fields. Though both
contributions originate from different physical mechanisms and even
have different symmetry tensor characters, their values could be
comparable in general case and should be taken into account on the
same footing.

The influences of a magnetic field and a stress on the DS of the FM+AFM
system are also different. The external magnetic field $\mathbf{H}$
discerns the domains with opposite orientations of the macroscopic
magnetization $\mathbf{M_{\rm F}}$ of FM (or FM component) and also
the domains with the perpendicular orientation of the AFM vector
$\mathbf{L}$, as follows from the standard expression for the Zeeman
energy\footnote{We distinguish a magnetization of the FM component
$\mathbf{M_{\rm F}}$ from a noncompensated magnetization
$\mathbf{M}$ of AFM. The later results from the field-induced
canting of the magnetic sublattices of AFM. } (per unit volume):
\begin{equation}\label{Zeeman_standard}
  w_{\rm
  Zee}=-\langle\mathbf{M}^{\rm{macro}}\rangle\mathbf{H}-\frac{\chi}{2}\langle[\mathbf{L}\times\mathbf{H}]^2\rangle,
\end{equation}
where $\chi$ is the magnetic susceptibility of the AFM subsystem,
and the macroscopic magnetization $\mathbf{M}^{\rm{macro}}$ includes
a contribution from both FM and AFM subsystems, see below, Eq.~(6).
It should be stressed that the domains with $\mathbf{L}\|\mathbf{H}$
are always unfavorable in a magnetic field and move away under the
field action.

On the contrary, the mechanical stress (described by the symmetric tensor
$\sigma_{jk}$) has no effect on the FM component (unless
it produces the magnetostriction) but may discern the AFM domains with
the perpendicular orientation of $\mathbf{L},$ as can be deduced from
the expression for the magnetoelastic energy (where the elastic
isotropy is supposed):
\begin{equation}\label{elastic energy}
  w_{\rm
  elas}=-\frac{\lambda}{2c_{44}}\langle L_jL_k-\frac{\mathbf{L}^2}{3}\delta_{jk}
  \rangle\left(\sigma_{jk}-\frac{{\rm Tr}\phantom{!}\hat\sigma}{3}\delta_{jk}\right),
\end{equation}
where $\lambda$ is the magnetoelastic constant, $c_{jk}$ are the 
elastic moduli, and we have omitted the terms related to the
isotropic (hydrostatic) pressure as immaterial for the further
consideration.

If, for example,
a tensile/compressive stress is applied along the $\mathbf{n}\|\mathbf{H}$
axis, then the combined action of the magnetic field and the stress is
described by the term (obtained from (3) and (4)):
\begin{equation}\label{combined}
   w_{\rm
  {field}}=-\langle\mathbf{M}^{\rm{macro}}\rangle\mathbf{H}-\frac{1}{2}\left(\chi H^2+\frac{\lambda\sigma}{c_{44}}\right)\langle[\mathbf{L}\times\mathbf{n}]^2\rangle,
\end{equation}
where we introduced the value of stress $\sigma\equiv{\rm
Tr}\phantom{!}\hat\sigma/3-\mathbf{n}\hat\sigma\mathbf{n}$.

It is quite obvious from Eq.~(5) that the action of a magnetic field
on the AFM subsystem is equivalent to the action of a stress with
fixed sign. So, a combination of $\mathbf{H}$ and $\hat\sigma$ gives
possibility to enhance (if $\lambda\sigma>0$) or wipe out (if
$\lambda\sigma\propto -c_{44}\chi H^2$) the inequivalence of
different AFM domains. Thus, it follows from general considerations
that the combined application of a stress and a magnetic field opens
a way to act selectively on that or those types of domains and to
control macroscopic properties of the sample (such as the
magnetization, elongation, magnetoresistance, {\it etc}.).

In the next sections, we consider the typical effects produced by
two fields in a relatively simple but nontrivial  AFM+FM
multiferroic Sr$_{2}$Cu$_{3}$O$_{4}$Cl$_{2}$.

\section{Model}
  The crystal structure of high-temperature superconducting cuprates Sr$_{2}$Cu$_{3}$O$_{4}$Cl$_{2}$ and
Ba$_{2}$Cu$_{3}$O$_{4}$Cl$_{2}$ consists of Cu$_{3}$O$_{4}$ planes
separated by spacer layers of SrCl or BaCl [15, 17]. Two types of
magnetic ions, CuI and CuII (see Fig. 1) form two interpenetrating
square lattices within Cu$_{3}$O$_{4}$ planes.

\begin{figure}
\includegraphics[width=\column]{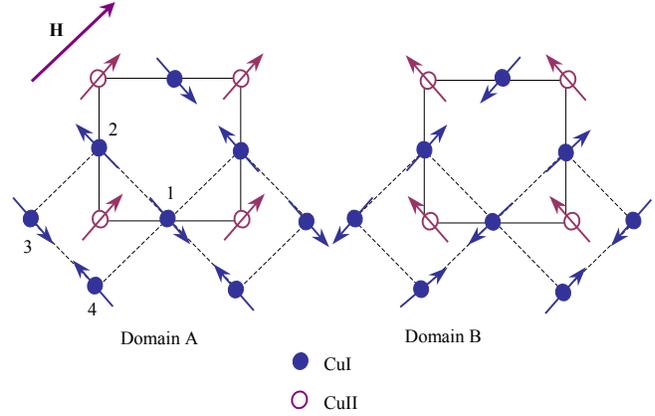}
\caption{ (Color online) Magnetic structure of Cu$_{3}$O$_{4}$ layer
in two  different  configurations (domains). Magnetic field is
parallel to $\langle  110\rangle$.  Two types of magnetic ions are
represented with filled and hollow circles.  FM ordered moments of
CuII could be ({\it a}) parallel (domain A) or ({\it b})
perpendicular (domain B) to the applied magnetic  field. Small
canting of the CuI spins induced by the external magnetic  field is
ignored }
\end{figure}

Within the temperature interval $T_{\rm II}\approx 40$~K$\le T \le
T_{\rm I}\approx 380$~K, the ions of the first type (CuI) are AFM
ordered, while the ions of the second type (CuII) bear a small but
nonzero FM moment\footnote{According to Ref. [18], the FM moments at
CuII ions result from the anisotropic ``pseudodipolar'' interactions
between CuI and CuII.}. According to the experiments [18], the mutual
orientation of CuI and CuII moments depends upon the direction of the
external magnetic field and can be either perpendicular or parallel.
In other words, the magnetic structure consists of two weakly
coupled subsystems, namely, the AFM subsystem localized on CuI ions and the FM
one localized on CuII ions. The FM subsystem is unambiguously
described by the magnetization vector $\mathbf{M}_{\rm F}$,  and the
AFM subsystem is described by two vectors: AFM vector
$\mathbf{L}=(\mathbf{S}_1-\mathbf{S}_2+\mathbf{S}_3-\mathbf{S}_4)/4$
and ferromagnetic vector $\mathbf{M}=\sum_j\mathbf{S}_j/4$
(numeration of CuI sites is shown in Fig. 1). The macroscopic
magnetization $\mathbf{M}^{\rm{macro}}$ (per unit volume) is defined
as a sum of FM and AFM magnetizations as follows:
\begin{equation}\label{macroscopic_ magnetization}
\mathbf{M}^{\rm{macro}}=\mathbf{M}_{\rm F}+\mathbf{M}.
\end{equation}

\begin{figure}
\includegraphics[width=\column]{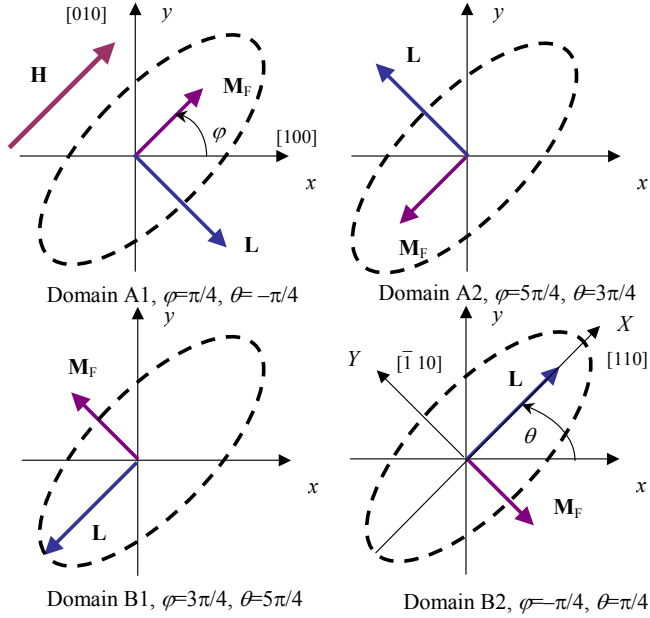}
\caption{ (Color online) Four types of magnetic domains. Axes $x$
and $y$ are parallel to $\langle 100\rangle$ crystal directions.
  The external magnetic field $\mathbf{H}\|[110]$ (if any). Types A
  and B have different orientations of the AFM vector, types 1 and 2
  correspond to opposite directions of the FM vector $\mathbf{M}_{\rm F}$.
Ellipse (dash line) images the supposed shape of the sample and its
orientation (axes $X$, $Y$) with respect to crystal axes }
\end{figure}

In the absence of an external field, the FM moments at CuII sites are
oriented along $\langle 110\rangle$ crystal directions perpendicular
to the staggered magnetizations of the AFM subsystem, as shown in
Fig.~1. Due to the tetragonal symmetry of the crystal (space group
$I4/mmm$), an equilibrium magnetic structure can be realized in four
types of equivalent domains, as shown in Figs.~1 and 2. Domains of
type A and B could be thought of as AFM domains, because they
correspond to different orientations of the $\mathbf{L}$ vector and thus
are sensitive to the orientation of the magnetic field $\mathbf{H}$ or
the stress $\hat\sigma$ with respect to the crystal axes (see Fig.~2).
Types A1 and A2 (and, respectively, B1 and B2) are FM domains;
they have opposite directions of the $\mathbf{M}_{\rm F}$ vector and
could be removed from the sample by $\mathbf{H}\|\mathbf{M}_{\rm
F}$.

If the external (magnetic and stress) fields are applied within the $xy$
(and, equivalently, $XY$) plane and their values are small enough as
compared with the exchange interactions between different Cu sites,
the magnetic structure of Sr$_{2}$Cu$_{3}$O$_{4}$Cl$_{2}$ can be
unambiguously described with two angle variables, as shown in Fig. 2:
\[ L_x=M_0\cos\theta,\quad L_y=M_0\sin\theta;\]
\begin{equation}\label{angles_parametrization}
M_{{\rm F}x}=m_{\rm F}M_0\cos\varphi,\quad M_{{\rm
  F}y}=m_{\rm F}M_0\sin\varphi.
\end{equation}
Here, $m_{\rm F}$(=$10^{-3}$ for Sr$_{2}$Cu$_{3}$O$_{4}$Cl$_{2}$
[18]) is a dimensionless constant that represents the ratio between
the spin moments localized on CuII and CuI sites. We have also taken
into account that, far below the N{\'e}el temperature, the
values of sublattice magnetizations $M_0=27.4$~Gs (that corresponds
to the spin $s=1/2$ per CuI site) and $M_{\rm F}$ are saturated and
constant.

The phenomenological description of the DS is based on the analysis of
the free energy potential $\Phi$ of the sample. We consider
four constituents of $\Phi$: magnetic $\Phi_{\rm{mag}}$, with
account of magnetoelastic interactions; shape-dependent stray
(demagnetizing) Eq. (1) and  destressing Eq. (2) energies, and the
energy $\Phi_{\rm{field}}\equiv V w_{\rm{field}}$ (see Eq. (5)) of
the external fields, as explained above:
\begin{equation}
\label{free_energy_1} \Phi = \Phi_{\rm{mag}} + \Phi_{\rm{stray}} +
\Phi_{\rm{dest}}+\Phi_{\rm{field}}.
\end{equation}
The magnetic energy of Sr$_{2}$Cu$_{3}$O$_{4}$Cl$_{2}$ crystal in the mean
field approximation is well established [9, 16, 18] and was analyzed
in details, along with $\Phi_{\rm{stray}}$  and $\Phi_{\rm{dest}}$,
in [19]. Here, we start directly from the simplified expression for
the specific potential $\phi\equiv \Phi/V$ expressed through the
angles $\theta$ and $\varphi$:
\[ \phi=4J_{\rm{pd}}m_{\rm F}\langle \cos(\theta+\varphi)\rangle+K_{\|}\langle
  \cos4\theta\rangle -\]
\[-\frac{J^2_{\rm{av}}}{8J_0}m^2_{\rm
  F}\langle\cos2(\theta-\varphi)\rangle+ \left(\!\frac{H^2}{32J_0}+\frac{\lambda M_0\sigma}{c_{44}}\!\right)
  \langle\cos2(\theta-\psi)\rangle-\]
\[-m_{\rm
F}H\left[\left(1-\frac{J_{\rm{av}}}{8J_0}\right)\langle\cos(\varphi-\psi)
\rangle+\right.\]
\[\left.
+\frac{J_{\rm{av}}}{8J_0}\langle\cos(2\theta-\psi-\varphi)\rangle
\right] -\]
\[-N^{\rm des}_{\rm 2an}\langle \cos2(\theta-\psi)\rangle
  +\frac{1}{2}M_0m_{\rm
F}^2\left[N^{\rm{dm}}_a\langle\cos(\varphi-\psi)\rangle^2+\right.\]
\[\left. +N^{\rm{dm}}_b\langle\sin(\varphi-\psi)\rangle^2\right]
+N^{\rm des}\langle\cos2(\theta-\psi)\rangle^2+\]
\begin{equation}\label{magnetic_energy_2}
 +\Delta N^{\rm
des}\langle\sin2(\theta-\psi)\rangle^2.
\end{equation}

In (9), $\psi$ is an angle between the unit vector
$\mathbf{n}\|\mathbf{H}$ and the $x$-axis, a mechanical stress is applied
along $\mathbf{n}$ as described above, $N^{\rm{dm}}_{a,b}$ are the
components of the demagnetization, and $N^{\rm des}$, $\Delta N^{\rm
des}$ are those of the destressing tensor which are important for the chosen
geometry of the sample (thin pillar of the thickness $c$ with an
elliptic cross-section, whose principal axes $X$ and $Y$ are
parallel to $\langle 110\rangle$ directions within the Cu$_3$O$_4$
layers, see Fig. 2), and $N^{\rm des}_{\rm 2an}$ is a shape-induced
anisotropy of the AFM subsystem.
  The meaning and values
(in Oe) of phenomenological constants used in (9) are given in Table
1. Here and for the rest of the paper, we use the values in Oe
instead of energy units (say, $\phi\rightarrow \phi/M_0$, {\it
etc}.). For the rest of the paper, we assume that the magnetic field
and the stress are applied in parallel to one of the principal axes of the
sample that coincides with an easy axis for the AFM vector, namely,
$\mathbf{n}\|X$, (see Fig. 2), so $\psi=\pi/4$. This situation
corresponds to the experimental setup in Refs. [9, 16].

\begin{table}[b]
\noindent\caption{Parameters used in the free energy potential [Eq.
(9)]. Raw data in meV taken from [9, 17, 18] are converted to Oe}
\label{table_data} \vskip3mm\tabcolsep3.6pt

\noindent{\footnotesize\begin{tabular}{cccc} \hline
\multicolumn{1}{c}{\rule{0pt}{9pt}
 Parameter} &\multicolumn{1}{|c} {Meaning}
   & \multicolumn{1}{|c}{Value (Oe)}
      \\
\hline %
\rule{0pt}{9pt}$J_0$& CuI--CuI superexchange (in-plane)&$1.02\times
10^7$\\
 $J_{\rm av}$& isotropic pseudodipolar interaction
&$-9.4\times 10^5$\\  $J_{\rm pd}$& anisotropic pseudodipolar
interaction&$-2.1\times 10^3$\\
$K_{\|}$& in-plane anisotropy $\times 10^{-6}$&$7.8\times
10^{-2}$\\\hline
\end{tabular}
}
\end{table}

\section{Homogeneous States in the Presence of Two Fields}
Let us start from the analysis of possible homogeneous states that
could be realized in the thermodynamic limit (infinite sample, all
$N^{\rm{dm}}, N^{\rm des}=0$). In this case, the minimization of $\phi$
with respect to the magnetic variables $\theta$ and $\varphi$ gives rise
to four solutions labeled as A1,2 and B1,2 (see Fig. 2).
Corresponding equilibrium values at $H=0$ and $\sigma=0$ are
\[{\textrm{A1:}}\quad  \theta_{\rm A1}=-\pi/4, \quad \varphi_{\rm A1}=\phantom{3}\pi/4;\]
\[{\textrm{B1:}}\quad  \theta_{\rm B1}=\phantom{3}5\pi/4, \quad \varphi_{\rm
 B1}=3\pi/4; \]
\[{\textrm{A2:}}\quad   \theta_{\rm A2}=3\pi/4, \quad \varphi_{\rm
  A2}=5\pi/4;\]
\begin{equation}\label{equilibrium_domains}
{\textrm{B2:}}\quad   \theta_{\rm B2}=\pi/4, \quad \varphi_{\rm
B2}=-\pi/4.
\end{equation}

 It should be stressed that, in contrast to pure AFMs,
the configurations with $(\mathbf{M}_{\rm F}, \mathbf{L})$ and
$(\mathbf{M}_{\rm F}, -\mathbf{L})$ are nonequivalent, due to
anisotropic pseudodipolar interactions (described by the constant
$J_{\rm pd}$).

\begin{figure}
\includegraphics[width=\column]{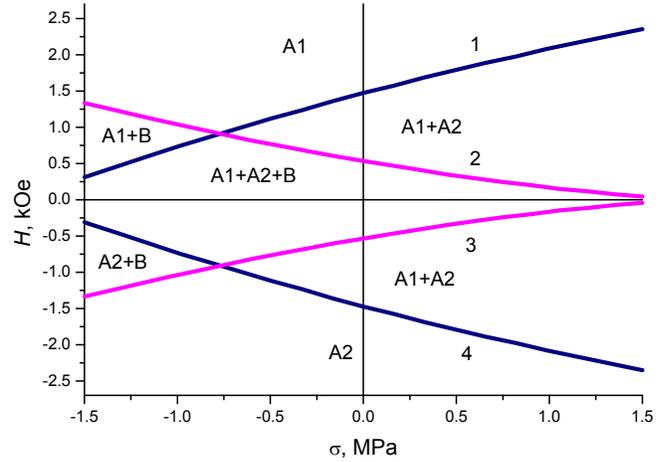}
\caption{ (Color online) Stability ranges of different
  homogeneous states in the presence of a field $H$ and a stress
  $\sigma$. Solid lines bound the stability ranges of A2 (line {\it 1}),
  B (lines {\it 2, 3}) and A1 (line {\it 4})  states }
\end{figure}

In the absence of a field, states A1,2 and B1,2 have the same energy and
are equivalent. Stress and magnetic fields remove the degeneracy of
these states, as can be seen from the phase diagram shown in Fig. 3.
Solid lines in Fig. 3 represent the stability ranges of different
states\footnote{~Since neither field nor stress in the accepted geometry
discern between B1 and B2 states, we consider both states as
B.} calculated by the direct minimization of potential (9). To
calculate the stress values, we used an estimation [19] for the
spontaneous strain $u_0=\lambda M_0^2/c_{44}\approx 10^{-6}$.

Lines {\it 2} and {\it 3} correspond to the step-like (spin-flop)
transition B1,2$\rightarrow$ A1 (line {\it 2}) and A2 (line {\it 3})
accompanied by the 90$^\circ$ rotation of the magnetic vectors
$\mathbf{L}$ and $\mathbf{M}_{\rm F}$. Within the regions that lie
below line {\it 1} and above line {\it 2} (and below line {\it 3}
and above line {\it 4}), potential (9) has only two minima that
correspond to states A1 and A2 with opposite orientations of the
magnetic vectors. Such a ``FM'' state can be induced solely by a
magnetic field. The effect of a stress reveals itself in the
appearance of the two- or three-state regions (above line {\it 1}
(3) and below line {\it 2} (4)), where equilibrium states A1 and B
(and A2 and B) have the perpendicular orientation of the magnetic
vectors. It is also worth to note the quite general result mentioned
in Sec.~2: the application of a stress can either enlarge or
diminish (depending on the sign of $\sigma$) the stability regions
of homogeneous states and, in that way, invert the order of phase
transitions between different states.

\begin{figure}
\includegraphics[width=\column]{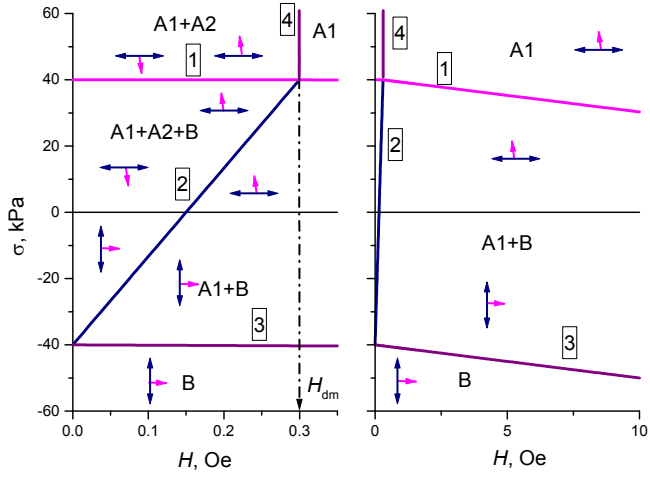}
\caption{ (Color online) $\sigma-H$ phase diagram of inhomogeneous
states for multiferroic
  Sr$_{2}$Cu$_{3}$O$_{4}$Cl$_{2}$. Left and right panels
  correspond to different field scales (below and above $H_{\rm
  dm}$). Solid lines are calculated from the condition of
  $\xi_j=0$, the composition of equilibrium DS is represented
  schematically in each region.
  Double-head and single-head arrows show orientations of the AFM and FM vectors,
  respectively. $\xi^{(0)}=0$ for simplicity }
\end{figure}

Thus, the combination of a stress and a field opens a way to
eliminate any state\footnote{~States B1 and B2 could be
discriminated by a small misalignment of the magnetic field from
[110] direction, as it was done, e.g., in Ref. [17].} and, as a
result, to produce single-domain states A1 (above lines {\it 1} and
{\it 2}) or A2 (below lines {\it 3} and {\it 4}) and any combination
of two- and three-domain states.

In the next sections, we consider the behavior of different
multidomain states which appear in the finite-size sample under the
application of a stress and a magnetic field.

\section{Stability Diagram of a Multidomain State}

In a finite-size sample, the equilibrium distribution of the vectors
$\mathbf{L}$ and $\mathbf{M}_{\rm F}$ is, in principle,
inhomogeneous due to boundary conditions on the sample surface. We
consider the multidomain states that consist of homogeneously
ordered regions (domains) separated by thin domain walls.
No inhomogeneous distribution of $\mathbf{L}$ and $\mathbf{M}_{\rm F}$
within the walls is taken into account. However, we assume that
the domain walls can move under ponderomotive, field-induced forces,
so that the DS can reach an equilibrium, after the field is applied. The
DS is thus unambiguously described by a set of angle variables
$\{\theta_j,\varphi_j\}$ ($j$~=~A1,~A2,~B1,~B2) and fractions
$\xi_j$ of domains of the $j$-type. Obviously, $\sum \xi_j$=1. The
number of free variables in the energy potential (9) depends upon
the types of movable domain walls present in the sample. The equilibrium
DS in the presence of an external field is then found from the
condition for the minimum of $\phi$ with respect to independent
variables.

In this section, we consider the case where all four types of
domains can freely grow or diminish in size (say, in a virgin sample
that initially contains domains of all types). If a stress $\sigma$
and a magnetic field $H$ are rather small (region A1~+~A2~+~B in
Fig. 4), both fields are screened by an appropriate domain
configuration, and the effective fields inside the sample are zero.
Formally, this means that $\xi_j$ are the only free variables. The
equilibrium values of the magnetic variables in this case are given
by Eq.~(10), and the domain fractions depend on the magnetic field
and the stress as follows:
\[\xi_{\rm A1,A2}=\frac{1}{4}\left[1-\xi^{(0)}\pm\frac{2H}{H_{\rm
dm}}+\left(\frac{H}{H_{\rm
  des}}\right)^2+\frac{\sigma}{\sigma_0}\right],\]
\begin{equation}\label{domain_fraction_4dom}
\xi_{\rm B1}=\xi_{\rm
B2}=\frac{1}{4}\left[1+\xi^{(0)}-\left(\frac{H}{H_{\rm
  {des}}}\right)^2-\frac{\sigma}{\sigma_0}\right],
\end{equation}
where we have introduced the notations
\begin{equation}\label{demag_field-def}
H_{\rm{dm}}\equiv m_{\rm F}N^{\rm{dm}}_aM_0, \quad H_{\rm
des}\equiv 8\sqrt{J_0N^{\rm des}},
\end{equation}  
\begin{equation}\label{zero_field_fraction}
\sigma_0\equiv \frac{2N^{\rm des}c_{44}}{\lambda M_0^2},\quad
\xi^{(0)}\equiv \frac{N^{\rm des}_{\rm 2an}}{N^{\rm des}}.
\end{equation}

Physically, $H_{\rm dm}$ is the FM demagnetizing field calculated as
if an AFM subsystem is absent. In an analogous way, $H_{\rm des}$
and $\sigma_0$ could be treated as the destressing field in the
absence of FM ordering expressed in the magnetic and stress
equivalents, respectively.
 The shape-dependent value $\xi^{(0)}$ represents the disbalance between type A and type B
  domain fractions
in the absence of a field. For the sample under consideration (see
Table~2), $\xi^{(0)}=0.22$ [19].

\begin{table}[b]
\noindent \caption{Parameters used in numerical simulations. The
source of data (experimental or calculated) is specified in the last
column. Destressing parameters correspond to \boldmath$T=120$~K.
\label{table_calc}}\vskip3mm\tabcolsep3.9pt \noindent{\footnotesize
\begin{tabular}{lccc}
\hline%
 \multicolumn{1}{c}{\rule{0pt}{9pt}Parameter}
 & \multicolumn{1}{|c}{Meaning}
 & \multicolumn{1}{|c}{Value}
 & \multicolumn{1}{|c}{Ref.}\\
 \hline%
\rule{0pt}{9pt}
 $a\times b\times c$& Sample
size&$7\times2\times0.5$~mm$^3$&[17]\\  $M_0$& Sublatt.
magnetization&27.4 Gs &[16]
\\$m_{\rm F}$& $M_{\rm F}/M_0$&$7\times 10^{-4}$&[16]
\\ $M_{\rm F}$& Satur. magnetization &$7\times
10^{-3}$~emu/g&[17]
\\  $\xi^{(0)}$ & Shape-induced bias
& 0.22 &[19]\\  $u_0$ & Spont. strain & $10^{-6}$& [19] \\ $H_{\rm
dm}$ &
Demagnetization field & 0.3 Oe & Eq.(\ref{demag_field-def})\\
$N^{\rm des}$ & Destressing
const. & $7$~mOe& [19] \\

 $H_{\rm des}$ & Destressing
field&$2.1$~kOe&[19]\\
  \hline\end{tabular}}
\end{table}

As seen from Eq.~(12), the value of magnetic destressing field
is enhanced due to the exchange interactions (constant $J_0$). On the
contrary, the demagnetizing field is weakened due to a small FM moment
($m_{\rm F}\ll 1$). So, the
demagnetizing field in the crystal under consideration is much smaller than the destressing ones,
$H_{\rm dm}\ll H_{\rm
  des}$ (see Table 2). However, for the chosen
  geometry of the sample, the
  demagnetizing and destressing effects are of the same order of
  value, as will be shown below.

The analysis of Eqs. (11) shows that the type of favorable domain
depends upon the sign\footnote{~The sign of $\sigma_0\propto\lambda$
is material-dependent and, in the case under consideration, is
unknown.} of $\sigma/\sigma_0$. If $\sigma/\sigma_0>0$, domains of B
type are unfavorable and disappear ($\xi_{\rm B}=0$), when (see line
{\it 1} in Fig. 4))
\begin{equation}\label{critical_B}
  \sigma_\mathrm{B}=\sigma_0\left[1+\xi^{(0)}-\left(\frac{H}{H_{\rm
  des}}\right)^2\right].
\end{equation}
A negative stress $\sigma/\sigma_0<0$ at $H=0$ can remove domains of
A2 type; respectively, $\xi_{\rm A1}=0$ at (line {\it 2} in Fig.~4))
\begin{equation}\label{critical_a2}
  \sigma_{\mathrm{A2}}=\sigma_0\left[\xi^{(0)}-1+\frac{2H}{H_{\rm{
  dm}}}-\left(\frac{H}{H_{\rm
  des}}\right)^2\right].
\end{equation}
Lines {\it 1} and {\it 2} in Fig. 4 bound the four-domain region and
intersect at the point $H=H_{\rm
  dm}$, $\sigma\approx\sigma_0\left(1+\xi^{(0)}\right)$. The two-
  domain state A1~+~A2 consists of FM domains and is
  stable in the region bounded by lines {\it 1} and {\it 4} (the latter corresponds to $H=H_{\rm
  dm}$).

  Three-domain state A1+B1+B2 (or A1+B, as B1 and B2 are
  equivalent) is equilibrium in a wide region below lines {\it 2}
  (Eq. (15)) and
  {\it 1} (Eq. (14)).
  The corresponding domain structure can be treated as an AFM one,
  because there are domains with different orientations of the $\mathbf{L}$ vector that compete with one another.
  It is worth to note that the negative stress
  $\sigma/\sigma_0<0$ makes the domains of B-type favorable.
  The magnetization vector $\mathbf{M}_{\rm F}$ of such domains is oriented \emph{perpendicularly} to an external magnetic field
  $\mathbf{H}$, as can be seen from the expression for equilibrium
  fractions
\[\xi_{\mathrm{A1}}=\frac{1}{2}\left[1-\xi^{(0)}+\frac{m_{\rm F}H}{4N^{\rm{des}}}+\left(\frac{H}{H_{\rm
  des}}\right)^2+\frac{\sigma}{\sigma_0}\right],\]
\begin{equation}\label{domain_fraction_3dom}
\xi_{\mathrm{B1}}=\xi_{\rm
B2}=\frac{1}{4}\left[1+\xi^{(0)}-\frac{m_{\rm
F}H}{4N^{\rm{des}}}-\left(\frac{H}{H_{\rm
  des}}\right)^2-\frac{\sigma}{\sigma_0}\right].
\end{equation}
In other words, the action of a stress may compensate the action of
a field. A negative stress can even remove A1-type domains (with
$\mathbf{M}_{\rm F}\|\mathbf{H}$) from the sample. This takes place
at the critical line (line {\it 3} in Fig. 4):
\begin{equation}\label{critical_a1}
  \sigma_{\mathrm{A1}}=\sigma_0\left[\xi^{(0)}-1-\frac{m_{\rm F}H}{4N^{\rm{des}}}-\left(\frac{H}{H_{\rm
  des}}\right)^2\right].
\end{equation}
Two more things are also worth noting. First, the lines within the
discussed phase diagram of the multidomain state correspond to the
reversible (second-order type) transitions realized through the
motion of domain walls. In contrast, the irreversible step-like
switching between domains (related to the disappearance or
nucleation of a certain state) can occur while crossing the
stability lines within the phase diagram of the homogeneous state
(Fig.~3).

Second, the scale of the reversible phase diagram (Fig.~4) is
proportional to a small shape-dependent demagnetizing field
$H_{\rm{dm}}=0.3$~Oe and a destressing  stress $\sigma_0\propto
40$~kPa. The scale of the irreversible phase diagram (Fig.~3) is
defined by the intrinsic spin-flop field $H_{\rm{s-f}}\propto 1$~kOe
[19] and the characteristic stress scale $\propto
  1$~MPa and is at least two
orders of magnitude larger. This scale difference is due to the
different activation barriers and, hence, different kinetics for
the nucleation and the domain wall motion.

In the next section, we consider the behavior of the DS consisting of
two types of domains with different orientations of $\mathbf{L}$
vectors.

\section{Competition of Two Domains: Nontrivial Field Dependence of Strain}

Let us consider a sample which was preliminary monodomainized to
state A1 by the excursion into the region of high fields (above
lines {\it 1} and {\it 2} in Fig. 3). Suppose that the magnetic
field $H$ is then diminished at a fixed stress value, $\sigma$.
Which kind of the DS can be expected in this case? It was already
shown in [19] that, at $H>0$ and $\sigma=0,$ metastable homogeneous
state B has lower energy than homogeneous state A2. The same
relation is true for a negative stress, $\lambda\sigma<0$, as the
energy of state B in this case is diminished, and also takes place
in a rather wide range of positive stress values (between lines {\it
2} and {\it 3} in Fig. 3). So, the nucleation of domains of B type
is much more probable than the nucleation of A2-type domains. If, in
addition, there is a slight misalignment between the magnetic field
$\mathbf{H}$ and a crystal axis [110] that removes the degeneracy
between the B1 and B2 states, the DS of the sample is represented by
the domains of only two types, A1 and~B1.\looseness=2

\begin{figure}
\includegraphics[width=\column]{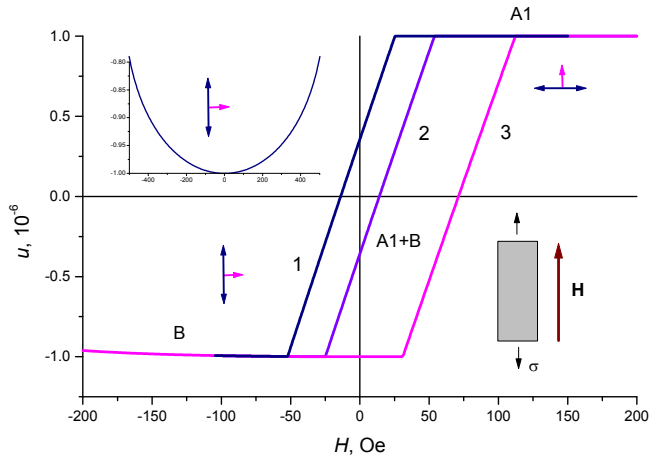}
\caption{ (Color online) Field dependence of the macroscopic strain
  $u(\mathbf{H})$. The sample contains domains of two types, A1 and B.
  Lines 1, 2 and 3 correspond to $\sigma=0.4$,
  $-0.4,$ and $-2$~kPa, respectively. Inset shows the same field
  dependence for a single domain (B-type) sample. Stress and
  magnetic field are applied as shown schematically in the right
  bottom angle }
\end{figure}

 To illustrate the peculiarities of macroscopic properties of such two-domain
 state, we calculated the dependence of the macroscopic
 magnetization, $\mathbf{M}^{\rm{macro}}$, and the elongation,
$u\equiv\langle \mathbf{n}\hat u\mathbf{n}-\mathrm{Tr}\hat
 u/2\rangle$ (where $\hat u$ is the strain tensor), on the magnetic field
 $\mathbf{H}$ at a fixed external stress.
Equilibrium values of the variables $\{\theta_j,\varphi_j,\xi_j\}$,
$j=A1,B1$, in this case were calculated by the numerical
minimization of potential (9), by using the data from Table 2 with
limitations $\xi_{\mathrm{A2}}=\xi_{\mathrm{B2}}=0$. The
corresponding dependences are given in Figs. 5 and~6.\looseness=1

The particular feature of the dependence $u(\mathbf{H})$ in Fig.~5
is its linearity which is due to the presence of A1 domain with
small FM moment. Really, in the single domain A1, the spontaneous
strain is constant. In the single domain B $u(\mathbf{H})$ shows the
quadratic, symmetric (with respect $\mathbf{H}\rightarrow
-\mathbf{H}$) behavior (inset in Fig. 5) that agrees well with the
symmetry of the strain tensor with respect to the time inversion
(and the corresponding inversion of $\mathbf{H}$). However, in the
two-domain case, it is the domain fraction $\xi$ that depends almost
linearly on $H,$ thus leading to a nontrivial and unusual field
dependence of the strain.\looseness=1

It can be also seen from Fig. 5 that the external stress induces a
bias (shift) of the $u(\mathbf{H})$ dependence, which is due to
the nonequivalence of domains A1 and B in the presence of a
stress. The same bias is induced into magnetization curves (see
Fig. 6). The value of bias is controlled by the stress. In thin
films, such a bias can be introduced into the sample either by an
external stress or by a properly chosen misfit between the film
and the substrate.

It is important to emphasize the role of a destressing energy in the
field-induced  behavior of the strain and the magnetization. In the case
under consideration, the demagnetizing (stray) energy is too small to
ensure the formation of metastable B-type domains. Really,
the demagnetization field $H_{\mathrm{dm}}=0.3$~Oe is much smaller than
the characteristic value of magnetic field, at which the appearance
of B-type domains is favorable, $H_{\mathrm{cr}}\propto
N^{\mathrm{des}}/m_{\rm F}\propto 100$~Oe [19] (see also Figs.~5 and 6).
So, the magnetic field-induced switching between A1-type and B1-type
domains is possible only due to destressing effects.\looseness=1

\begin{figure}
\includegraphics[width=\column]{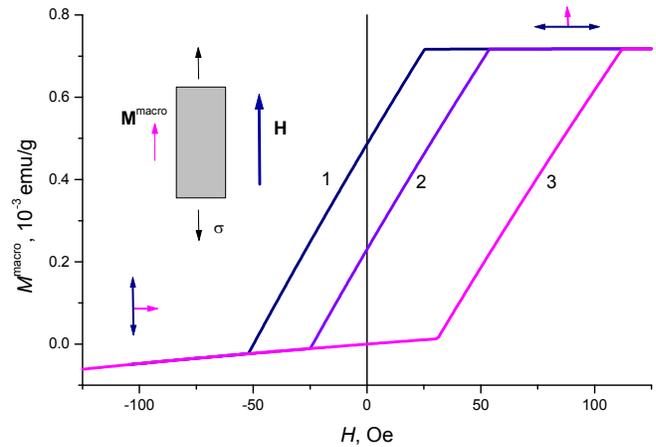}
\caption{ (Color online) Field dependence of the macroscopic
  magnetization
  $M^{\mathrm{macro}}(\mathbf{H})$. The sample contains domains of two types, A1 and B.
  Lines 1, 2 and 3 correspond to $\sigma=0.4$,  $-0.4,$ and $-2$~kPa, respectively
   \label{fig_magnetization_vs_field+stress} }
\end{figure}

\section{Conclusions}
We have considered the behavior of the AFM+FM multiferroic under the
action of two external fields of different nature. Our calculations
show that such a combination of fields makes it possible to separate
the field influence on the different coexisting order parameters.
This, in turn, opens a way to control the DS and macroscopic
properties and to produce the states with any desirable types of
domains.

The simultaneous application of two fields can also increase
the susceptibility of a sample to one of the fields, like it is the
case with the magnetic susceptibility of a strain. The range
 of the field values, in which the system is sensible to external
 fields, can be controlled by the appropriate choice of the sample shape
 and the corresponding shape-induced ``de-'' fields (demagnetizing,
 depolarizing, destressing, {\it etc.}).\looseness=1

For the present work, we have chosen the multiferroic with a rather
simple, tetragonal structure. However, it would be interesting to
generalize the proposed approach to multiferroics with ferroelectric
and magnetic order parameters that show simultaneously electro- and
magnetostriction (like PZT-PFW systems [20]) or/and have
incommensurate AFM structure (like BiFeO$_3$) that can be realized
in more than four types of domains. We believe that three types of
possible control fields (electric, magnetic, and stress) in the
first case and a large number of possible equilibrium domains in the
second case can give rise to some new peculiarities in the
macroscopic behavior of multiferroics.\looseness=1

\vskip3mm The authors acknowledge the financial support from the
Division of Physics and Astronomy of the National Academy of
Sciences of Ukraine in the framework of Special Program for
Fundamental Research. The work was partially supported by the grant
from the Ministry of Education and Science of Ukraine.\looseness=1

\rezume{ТЕОРЕТИЧНИЙ ОПИС\\ ФАЗОВОЇ ДІАГРАМИ МУЛЬТИФЕРОЇКІВ\\
СКІНЧЕННИХ РОЗМІРІВ У ЗМІННИХ МЕХАНІЧНЕ\\ НАПРУЖЕННЯ--МАГНІТНЕ~~
ПОЛЕ:~~~ КОНКУРЕНЦІЯ\\ МІЖ ФЕРО- ТА АНТИФЕРОМАГНІТИМИ ДОМЕНАМИ}{О.В.
Гомонай, Є. Г. Корнієнко, В.М. Локтєв} {Макроскопічними
властивостями мультифероїків, тобто речовин, в яких співіснують два
типи впорядкування, можна керувати за допомогою полів різної
природи. В даній роботі проаналізовано поведінку мультифероїків з
антиферо- (АФМ) та феромагнітним (ФМ) впорядкуванням під дією
зовнішнього магнітного поля та механічного напруження. Комбінація
цих двох полів дозволяє отримати макроскопічні стани з різною
доменною структурою. Отриманий  таким способом дводоменний стан
демонструє нетипову для АФМ лінійну залежність макроскопічної
деформації від зовнішнього магнітного поля. Мале, але ненульове
механічне напруження приводить також до зсуву кривої залежності
макроскопічної намагнічуваності від магнітного поля (ефективному
``підмагнічуванню'').}

\end{document}